\documentclass[aps,pra,twocolumn,showpacs,preprintnumbers,amsmath,amssymb,footinbib]{revtex4}
\usepackage{mathtools}

\DeclareMathOperator{\sinc}{sinc}
\DeclareMathAlphabet{\mathcalligra}{T1}{calligra}{m}{n}
\DeclareMathAlphabet{\mathpzc}{OT1}{pzc}{m}{it}

\usepackage{graphicx,epsfig}
\usepackage{bm}
\usepackage{dcolumn}
\usepackage{xcolor}
\usepackage[breaklinks=true,colorlinks,citecolor=blue,linkcolor=blue,urlcolor=blue]{hyperref}

\def\ajp#1#2#3{Am. J. Phys. {\bf #1}, #2 (#3)}
\def\ejp#1#2#3{Eur. J. Phys. {\bf #1}, #2 (#3)}

\def\noi{\noindent}
\def\bc{\begin{center}}
\def\ec{\end{center}}
\newcommand{\bea}{\begin{equation}}
\newcommand{\eea}{\end{equation}\noi}
\newcommand{\ber}{\begin{eqnarray}}
\newcommand{\eer}{\end{eqnarray}\noi}
\begin{document}
\title{Explicit derivation of the Fraunhofer diffraction formula for oblique incidence}

\author{Shyamal Biswas$^{1}$}\email{sbsp [at] uohyd.ac.in}
\author{Rhitabrata Bhattacharyya$^{2}$}
\author{Saugata Bhattacharyya$^{3}$}
\affiliation{$^{1}$School of Physics, University of Hyderabad, C.R. Rao Road, Gachibowli, Hyderabad-500046, India \\
$^{2}$Central GST \& Central Excise, Kolkata North Commissionerate, Kolkata-700107, India\\ 
$^{3}$Department of Physics, Vidyasagar College, 39 Sankar Ghosh Lane, Kolkata-700006, India
}

\date{\today}

\begin{abstract}
We have analytically explored the Rayleigh-Sommerfeld scalar diffraction for oblique incidence. We have explicitly derived the Fraunhofer diffraction formulae for oblique incidence of plane scalar wave on various apertures, such as single-slit, circular aperture, and diffraction grating. Such derivations in the background of Rayleigh-Sommerfeld scalar diffraction theory would be important for an undergraduate course on optics.
\end{abstract}


\maketitle    

\section{Introduction}
The Fraunhofer diffraction formula \cite{Born,Goodman,Lahiri} is certainly taught in an undergraduate course on optics. Derivation of the Fraunhofer diffraction often assumes normal incidence of scalar wave on apertures such as single-slit, double-slit, diffraction grating, circular aperture, \textit{etc} \cite{Goodman,Lahiri}. A natural question may be asked in a class about the generalization of the Fraunhofer diffraction formula for the oblique incidence. The generalized Fraunhofer diffraction formula (i.e. the grating equation for oblique incidence) exists in the literature \cite{Born2} for a diffraction grating without its explicit derivation. This makes it difficult for undergraduate students to follow the topic from the mathematical point of view. The generalization for an arbitrary aperture, in our opinion, is a nontrivial task because it requires integration of the space-dependent phase of the scalar wave obliquely incident on the aperture. This motivates us to explicitly derive the generalized Fraunhofer diffraction formulae for different apertures. Apparently, this excursion will not only engage the young minds to use mathematical techniques but also will greatly facilitate a deeper appreciation of the problem.  

Diffraction of electromagnetic wave by an aperture (or an obstacle) still continues to be a subject of great interest as an application of the Huygens-Fresnel principle in both the basic sciences and the applied sciences \cite{Sommerfeld2,Born,Goodman,Lucke,Lahiri,Bhattacharyya}. Proper analysis of the diffraction of electromagnetic wave though needs the vector diffraction theory \cite{Born}, Rayleigh-Sommerfeld scalar diffraction theory of the type-I (and also the type-II) \cite{Sommerfeld2} serves the purpose if the dimension of the aperture is larger than the wavelength of the scalar field incident on it. We are interested in deriving the Fraunhofer diffraction formula from the Rayleigh-Sommerfeld scalar diffraction theory of type-I because it is `manifestly' consistent \cite{Mukunda} and matches better with the experimental data \cite{Zurak} for the oblique incidence. Diffraction of the electromagnetic wave was already discussed in great detail for oblique incidence of both the vector field \cite{Vector-Oblique} and the scalar field \cite{Scalar-Oblique,Scalar-Oblique2}. Fraunhofer diffraction formula for the oblique incidence, however, has not been derived in this literature \cite{Vector-Oblique,Scalar-Oblique2} not even in the pedagogical one \cite{Scalar-Oblique}. Diffraction of the scalar field is quite common for the oblique incidence of a spherical wave \cite{Born,Goodman,Lucke}. The Fraunhofer diffraction formula, however, is not derived for spherical waves rather for plane waves. On the other hand, Fraunhofer diffraction formula for the oblique incidence has been stated with a qualitative derivation within a geometrical description for a diffraction grating in Ref.\cite{Born2}. Explicit derivation of the Fraunhofer diffraction formulae for oblique incidence on single-slit, diffraction grating, circular aperture, \textit{etc}, however, are not found in the literature. Hence we take up the discussion of the explicit derivation of the generalized Fraunhofer diffraction formulae. Such a project can be taken up within the scope of the undergraduate course on physics.   

Calculation in this article begins with the wave equation for the Rayleigh-Sommerfeld scalar diffraction \cite{Sommerfeld2}. We take far-field approximation of the Rayleigh-Sommerfeld integral (of the type-I) to get the Fraunhofer diffraction equation. We consider oblique incidence of a monochromatic plane scalar wave on the single-slit, circular aperture, and diffraction grating. Here-from we get Fraunhofer diffraction formulae for the oblique incidence on these apertures. Finally, we conclude. 

\section{Rayleigh-Sommerfeld scalar diffraction for oblique incidence}

\subsection{Exact result for oblique incidence}

\begin{figure}
\includegraphics[width=.98 \linewidth]{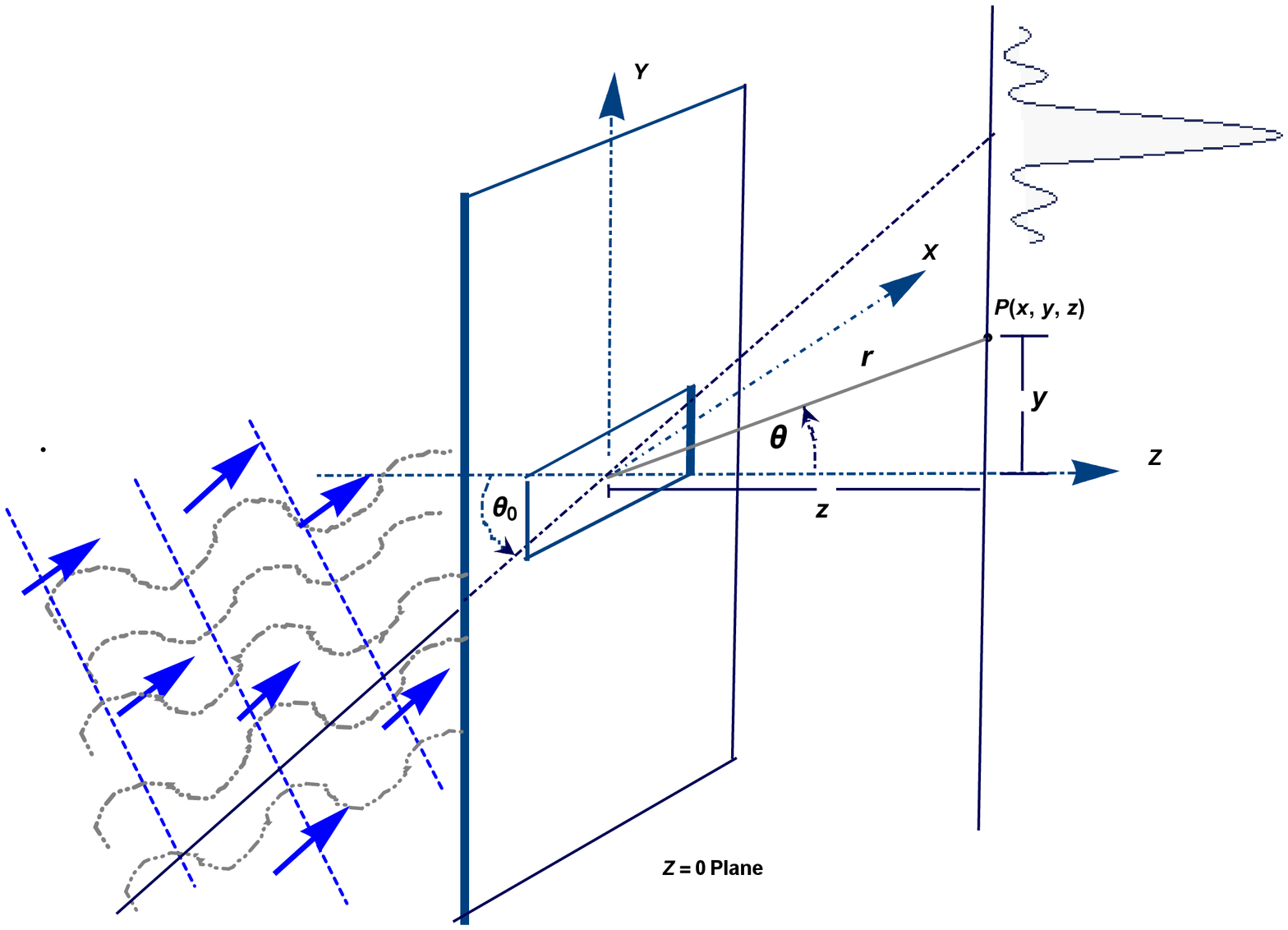}
\caption{Schematic diagram for diffraction by a rectangular aperture for oblique incidence.
\label{fig1}}
\end{figure}

Let us consider a monochromatic plane scalar wave incidents on an aperture $\mathcal{A}$ with the angle of incidence $\theta_0$ from the left half-space $z<0$ as shown in figure \ref{fig1}. Scalar diffraction incidentally takes place in the right-half space $z>0$ due to the aperture situated in the $x-y$ plane. Let us fix the origin of the Cartesian coordinate system to be at the middle of the aperture. Let a point on the aperture be given by $\vec{r}_{\perp_0}=x_0\hat{i}+y_0\hat{j}$.  The scalar field $\psi(\vec{r},t)$ at the position $\vec{r}=x\hat{i}+y\hat{j}+z\hat{k}$ and time $t$ satisfies the wave equation \cite{Born,Goodman,Lahiri}
\begin{eqnarray}\label{eq1}
\Big(\frac{1}{c^2}\frac{\partial^2}{\partial t^2}-\nabla^2 \Big)\psi(\vec{r},t)=0
\end{eqnarray}
where $c$ is the speed of light in the vacuum. The scalar field $\psi(\vec{r},t)$ is equated with one of the components of the electric field ($\vec{E}(\vec{r},t)$) or magnetic field ($\vec{B}(\vec{r},t)$) of the electromagnetic field \cite{Goodman,Bhattacharyya}. The scalar field can be written in the separable form $\psi(\vec{r},t)=U(\vec{r})\text{e}^{-i\omega t}$ for the monochromatic wave of angular frequency $\omega$. Eventually, the time-independent part of the scalar field satisfies the Helmholtz equation \cite{Born,Goodman,Lahiri}
\begin{eqnarray}\label{eq2}
(\nabla^2+k^2)U(x,y,z)=0
\end{eqnarray}
where $k=\omega/c$ is the wavenumber of the monochromatic wave. For $z>0$, the time-independent part of the scalar field can be expressed as the Rayleigh-Sommerfeld integral of the type-I as \cite{Sommerfeld}
\begin{eqnarray}\label{eq3}
U(x,y,z)=\frac{1}{4\pi}\iint_{\mathcal{A}}U(x_0,y_0,0)\bigg[\frac{\partial G_-(\vec{r}) }{\partial z_0}\bigg]_{z_0=0}\text{d}x_0\text{d}y_0
\end{eqnarray}
where $R_+=+[(x-x_0)^2+(y-y_0)^2+(z-z_0)^2]^{1/2}$, $R_-=+[(x-x_0)^2+(y-y_0)^2+(z+z_0)^2]^{1/2}$ and $G_-(\vec{r})=\frac{\text{e}^{ikR_+}}{R_+}-\frac{\text{e}^{ikR_-}}{R_-}$. Here $U(x_0,y_0,0)$ represents the time-independent part of the incident wave. For the oblique incidence of a monochromatic plane scalar wave, we have $U(x_0,y_0,0)=\text{e}^{ik\sin(\theta_0)y_0}$ where $\theta_0$ is the angle of incidence as shown in figure \ref{fig1}. The expression $U(x_0,y_0,0)=u_0\text{e}^{i\vec{k}\cdot\vec{r}_{\perp_0}}=u_0\text{e}^{ik\sin(\theta_0)y_0}$, where $u_0$ is a constant, is clearly evident from figure \ref{fig1} because the indecent wave vector $\vec{k}$ makes an angle $\pi/2-\theta_0$ with the $y$-axis in the aperture and $\pi/2$ with the $x$-axis in the aperture. Thus we recast Eqn. (\ref{eq3}) as
\begin{eqnarray}\label{eq4}
U(\vec{r})&=&-\frac{zu_0}{2\pi}\iint_{\mathcal{A}}\text{e}^{i[k\sin(\theta_0)y_0+kR(x,x_0,y,y_0,z)]}\times\nonumber\\&&\bigg[\frac{ik}{R^2(x,x_0,y,y_0,z)}-\frac{1}{R^3(x,x_0,y,y_0,z)}\bigg]\text{d}x_0\text{d}y_0\nonumber\\
\end{eqnarray}
where $R(x,x_0,y,y_0,z)=+[(x-x_0)^2+(y-y_0)^2+(z-0)^2]^{1/2}$. This result, however, is exact and is consistent with the Kirchhoff boundary conditions. The $y_0$-dependence of the phase in the above equation makes the integration over the secondary source points $\{x_0,y_0\}$ nontrivial even for the far-field case of the observation point $\vec{r}$. It should be mentioned in this regard that Eqn. (\ref{eq3}) where-from we get Eqn. (\ref{eq4}), can also be obtained by the integral transform formulation of the diffraction theory as described in Ref.\cite{Sherman}. 

\subsection{Far-field zone result for oblique incidence}
Let the wavelength of the incident wave be $\lambda=\frac{2\pi}{k}$ and the typical dimension of the aperture be $D$. By far-field zone we mean $\frac{D^2}{2r\lambda}\ll1$ and by the applicability of the scalar diffraction theory we mean $\lambda\ll D$ \cite{Born,Goodman,Lahiri}. The distance between a secondary source point $\vec{r}_{\perp_0}$ and an observation point $\vec{r}$ in the far-field zone can be approximated as
\begin{eqnarray}\label{eq5} 
R&=&+[(x-x_0)^2+(y-y_0)^2+z^2]^{1/2}\nonumber\\&\simeq&r\bigg[1-\frac{xx_0+yy_0}{r^2}\bigg].
\end{eqnarray}
With this form of $R$, we approximate Eqn. (\ref{eq4}) as 
\begin{eqnarray}\label{eq6}
U(\vec{r})\simeq-\frac{ikzu_0\text{e}^{ikr}}{2\pi r^2}\iint_{\mathcal{A}}\text{e}^{-ik[(xx_0+yy_0)/r-\sin(\theta_0)y_0]}\text{d}x_0\text{d}y_0\nonumber\\
\end{eqnarray}
for $\frac{D^2}{2r\lambda}\ll1$ and $\lambda\ll D$. Eqn. (\ref{eq6}) can be called as the Fraunhofer diffraction equation for the oblique incidence. The spatial part of the scalar field in Eqn. (\ref{eq6}) is nothing but the Fourier transform of the aperture function for the oblique incidence. The presence of the phase term $k\sin(\theta_0)y_0$ in the Fourier transform breaks the symmetry of the intensity pattern around $y=0$. Hence Eqn. (\ref{eq6}) is our key result for obtaining the Fraunhofer diffraction formula for oblique incidence on any 2-D aperture.

\subsubsection{Fraunhofer diffraction formula for oblique incidence on a single-slit}
For a rectangular aperture of length $a$ and breadth $b$, we have $-a/2<x_0<a/2$ and $-b/2<y_0<b/2$. Integrations over $x_0$ and $y_0$ in Eqn. (\ref{eq6}) result in
\begin{eqnarray}\label{eq7}
U(\vec{r})\simeq-\frac{iabzu_0\text{e}^{ikr}}{\lambda r^2}\sinc\bigg(\frac{\pi a x}{\lambda r}\bigg)\sinc\bigg(\frac{\pi b [y-r\sin(\theta_0)]}{\lambda r}\bigg)
\end{eqnarray}
for the rectangular aperture. According to figure \ref{fig1}, we have $y/r=\sin(\theta)$ (where $\theta$ is the angle of diffraction) and $z/r=\cos(\theta)$ for $x\rightarrow0$. Thus for the diffraction along the $y$-axis, we get the generalized Fraunhofer diffraction formula from Eqn. (\ref{eq7}) as 
\begin{eqnarray}\label{eq8}
b[\sin(\theta)-\sin(\theta_0)]=n\lambda
\end{eqnarray}
where $n=\pm1, \pm2, \pm3,...$ represent the location of the intensity ($\propto |U(\vec{r})|^2$) minima. The principal maximum of the intensity, of course, occurs at $\theta=\theta_0$. From the geometrical point of view, we can say that two parallel light-rays passing through two points separated by a distance $b$ on the single-slit, make a path difference $-b\sin(\theta_0)$ (for the angle of incidence $\theta_0$) while reaching the aperture and make another path difference $b\sin(\theta)$ (for the angle of diffraction $\theta$) while reaching the observation point after passing the aperture. These two rays altogether make $b[\sin(\theta)-\sin(\theta_0)]$ amount of path difference from their common source point to their observation point. Phase difference corresponding to this amount of path difference is $\frac{2\pi}{\lambda}b[\sin(\theta)-\sin(\theta_0)]$. Now we can safely say that if this amount of phase difference is $2\pi n$ (where $n=\pm1, \pm2, \pm3,...$), then for the intensity-minima, we have $\frac{2\pi}{\lambda}b[\sin(\theta)-\sin(\theta_0)]=2\pi n$ which is nothing but Eqn. (\ref{eq8}). However, the geometrical point of view does not clearly say why the distance between the two points on the single-slit would be $b$. It could be less than $b$ too. Geometrical point of view does not also clearly say why the intensity-minima occur for $b[\sin(\theta)-\sin(\theta_0)]=n\lambda$. It could be maxima too. So, the geometrical point of view is limited to explain the fringe pattern even for a single-slit diffraction. Hence we need a mathematical point of view with Eqn. (\ref{eq3}) as a starting point for the rigorous description of diffraction for oblique incidence even on a single-slit. However, is clear from Eqn. (\ref{eq8}) that the oblique incidence results in an extra $-2\pi b\sin(\theta_0)/\lambda$ amount of phase difference for shifting the principal maximum from $\theta=0$ to $\theta=\theta_0$.

\subsubsection{Fraunhofer diffraction formula for oblique incidence on a circular aperture}
Let us replace the rectangular aperture by a circular aperture of radius $a$ in figure \ref{fig1}. Appearance of the oblique incidence term $\sin(\theta_0)y_0$ in Eqn. (\ref{eq6}) breaks the radial symmetry of the scalar field around $\theta=0$. We evaluate the integrals in Eqn. (\ref{eq6}) for the circular aperture in circular polar coordinates ($r_0,\phi_0$) (where $x_0=r_0\cos(\phi_0)~\text{and}~y_0=r_0\sin(\phi_0)$) for $0<r_0\le a$ and $0\le\phi_0<2\pi$, as 
\begin{eqnarray}\label{eq9}
U(\vec{r})\simeq-\frac{2\pi a^2zu_0\text{e}^{ikr}}{\lambda r^2}\frac{\text{J}_1\big(\frac{2\pi a}{\lambda}\sqrt{\big[\frac{y}{r}-\sin(\theta_0)\big]^2+\big[\frac{x}{r}\big]^2}\big)}{\frac{2\pi a}{\lambda}\sqrt{\big[\frac{y}{r}-\sin(\theta_0)\big]^2+\big[\frac{x}{r}\big]^2}}
\end{eqnarray}
where $J_1$ represents the Bessel function of order $1$ of the first kind. It should be mentioned in this regard that vector generalization of Eqn. (\ref{eq9}) results from the vector Smythe-Kirchhoff formula and the generalized result is expected as a solution to problem number 10.12 of Ref.\cite{Jackson}. However, again we have $y/r=\sin(\theta)$ and $z/r=\cos(\theta)$ for $x\rightarrow0$. Thus for the diffraction along the $y$-axis, we get the generalized Fraunhofer diffraction formula from Eqn. (\ref{eq9}) as 
\begin{eqnarray}\label{eq10}
2a[\sin(\theta)-\sin(\theta_0)]=\frac{\gamma_{1,n}}{\pi}\lambda
\end{eqnarray}
where $\gamma_{1,n}$ is the $n$th zero of $J_1$ and $n=\pm1, \pm2, \pm3,...$. The Bessel zeros represent the location of the intensity minima. It is clear from Eqn. (\ref{eq10}) that the oblique incidence results in an extra $-4\pi a\sin(\theta_0)/\lambda$ amount of phase difference for shifting the principal maximum of the intensity ($\propto |U(\vec{r})|^2$) from $\theta=0$ to $\theta=\theta_0$. 

\subsubsection{Fraunhofer diffraction formula for oblique incidence on a grating}
Let us replace the rectangular aperture with a diffraction grating in figure \ref{fig1}. The diffraction grating is an array of $N$ single-slits each of which has length $a$ along the $x$-axis and breadth $b$ along the $y$-axis. Middle of the $j$th slit with respect to that of the middle one is separated by $y_j=jd$ where $j=0, \pm1, \pm2,..., \pm(N-1)/2$. We further take the approximation that $j^2d^2/r^2\ll1$. Thus Eqn. (\ref{eq6}) for the $j$th single-slit would be read as
\begin{eqnarray}\label{eq11}
U_j(\vec{r})&\simeq&-\frac{ikzu_0\text{e}^{ikr}}{2\pi r^2}\times\nonumber\\&&\iint_{\mathcal{A}_j}\text{e}^{-ik\big[(xx_0+y[y_0+jd])/r-\sin(\theta_0)[y_0+jd]\big]}\text{d}x_0\text{d}y_0\nonumber\\
\end{eqnarray}
where $\mathcal{A}_j$ represents aperture of the $j$th single-slit and $-a/2<x_0<a/2$ and $-b/2<y_0<b/2$. The time-independent part of the scalar field at the observation point $\vec{r}$ is obtained as a superposition of the contribution of the individual single-slits which differ in their contribution by their respective phases. Thus we get the time-independent part of the scalar field at the observation point $\vec{r}$ in the far-field zone as 
\begin{eqnarray}\label{eq12}
U(\vec{r})&\simeq&\sum_{-(N-1)/2}^{(N-1)/2}U_j(\vec{r})\nonumber\\&=&
-i\frac{abz}{\lambda r^2}u_0\text{e}^{ikr}\frac{\sin(\frac{\pi dN[y/r-\sin(\theta_0)]}{\lambda})}{\sin(\frac{\pi d[y/r-\sin(\theta_0)]}{\lambda})}\nonumber\\&&\times\sinc\bigg(\frac{\pi b[y/r-\sin(\theta_0)]}{\lambda}\bigg)\times\sinc\bigg(\frac{\pi ax/r}{\lambda}\bigg).~~~~~
\end{eqnarray}
It is clear from Eqn. (\ref{eq12}) that the principal maxima of the intensity ($\propto|U(\vec{r})|^2$) are appearing for
\begin{eqnarray}\label{eq13}
d[\sin(\theta)-\sin(\theta_0)]=n\lambda
\end{eqnarray}
where $\sin(\theta)=y/r$ for $x\rightarrow0$ and $n=0, \pm1, \pm2,...$. Eqn. (\ref{eq13}) is the generalized Fraunhofer diffraction formula (also called as grating equation) for the diffraction grating. This equation in not new and is in found in the literature \cite{Born2} without its explicit derivation. It should be mentioned in this regard that this formula remains unaltered for the oblique incidence (with $\theta$ as the angle of reflection) on a reflection grating \cite{Hecht,Dmitrieva}. Derivation of the generalized Fraunhofer diffraction formula for the reflection grating would be similar to that described in this subsection.  

\section{Conclusion}

To conclude, we have explicitly derived the Fraunhofer diffraction formulae for oblique incidence of plane scalar wave on various apertures, such as single-slit, circular aperture, and diffraction grating. Our results are essentially not new. However, integrations over the space-dependent phase of the scalar wave obliquely incident on the apertures make our derivations nontrivial. Such derivations in the background of Rayleigh-Sommerfeld scalar diffraction theory would be important for an undergraduate course on optics. Our article will be directly useful in the classroom.

Fraunhofer diffraction formula for oblique incidence is not unknown to physicists at least for the diffraction grating. Some textbooks assign this problem from the geometrical point of view to the end of relevant sections or chapters \cite{Born2,Hecht}. However, we have explained that the geometrical point of view is limited to well describe the fringe pattern for the Fraunhofer diffraction. Hence a rigorous mathematical description is needed, like the one presented here, to obtain the Fraunhofer diffraction formula for oblique incidence.

Calculation in the manuscript could have been started with Eqn. (\ref{eq6}) by generalizing Eqn. (4.38) of Ref.\cite{Goodman} for oblique incidence and large angle of diffraction. Such a starting with the Fourier optics formalism would be suitable for a special audience. Undergraduate students, however, may feel easy with the wave equation (Eqn. (\ref{eq1})) as the starting point for the rigorous derivation of the fringe pattern for the Fraunhofer diffraction. Hence we have started our calculation from the wave equation.

It is interesting to note that there is an apparent similarity of our work on the rectangular aperture with the work in Ref.\cite{Marvin} where Fraunhofer diffraction by a parallelogram-shaped aperture has been discussed both from a theoretical and an experimental point of view. The complement of the smaller interior angle of the parallelogram results in such similarity. Our work is, however, significantly different from the work of Ref.\cite{Marvin} because the only normal incidence of a plane scalar wave has been theoretically considered in Ref.\cite{Marvin}. On the other hand, we have considered the oblique incidence of the plane scalar waves for all the cases of the apertures.   

Other scalar diffraction theory, such as Fresnel-Kirchhoff theory \cite{Born}, would result in the same generalized Fraunhofer diffraction formulae at least for small angle of diffraction and small angle of incidence because the obliquity factor ($\cos(\theta)=\frac{z}{r}$ \cite{Goodman}) does not play any significant role in determining location of the intensity-maximum in such a case. However, we have not considered this theory for our analysis because it is not consistent with the boundary conditions \cite{Sommerfeld2}.

We have considered the number of single-slits to be odd for the diffraction grating. The Fraunhofer diffraction equation we have obtained in Eqn. (\ref{eq12}) would, however, be applicable for any number of single-slits. Hence the Fraunhofer diffraction formula we have obtained in Eqn. (\ref{eq13}) would be applicable for a double-slit too. Undergraduate students can directly derive the Fraunhofer diffraction formula for the oblique incidence on a double-slit by following the procedure described in this article.

Here we have discussed the scalar diffraction (i.e. the Fraunhofer diffraction) in the far-field zone. Scalar diffraction (i.e. the Fresnel diffraction) in the radiative near-field zone is also followed from the Helmholtz equation (i.e. Eqn. (\ref{eq2})). The behaviour of the diffracted field, in such a case, can be described in terms of the pseudo-plane waves and the pseudo-time in the Schr$\ddot{\text{o}}$dinger-like equation or in the paraxial equation \cite{Grella}.

\section*{Acknowledgement}
S. Biswas acknowledges partial financial support of the SERB, DST, Govt. of India under the EMEQ Scheme [No. EEQ/2019/000017].

\end{document}